# Spectroscopic parameters related to non bridging oxygen hole centers in amorphous-SiO$_2$

L. Vaccaro and M. Cannas*

*Dipartimento di Scienze Fisiche ed Astronomiche dell'Università di Palermo, INFM,*

*Via Archirafi 36, I-90123 Palermo, Italy*

B. Boizot

*Laboratoire de Solides Irradiés, CEA/DSM/DRECAM/, Palaiseau, France*

The relationship between the luminescence at 1.9 eV and the absorption bands at 2.0 eV and at 4.8 eV were investigated in a wide variety of synthetic silica samples exposed to different γ- and β-ray irradiation doses. We found that the intensities of these optical bands are linearly correlated in agreement with the model in which they are assigned to a single defect. This finding allows to determine spectroscopic parameters related to optical transitions efficiency: the oscillator strength of the 4.8 eV results ~200 times higher than that of the 2.0 eV; the 1.9 eV luminescence quantum yield under 4.8 eV excitation is lower (by a factor ~3) than that under 2.0 eV excitation. These results are consistent with the energetic level scheme, proposed in literature for non bridging oxygen hole center, and account for the excitation/luminescence pathways occurring after UV and visible absorption.

* Corresponding author: Phone: +390916234298; Fax: +390916162461; E-mail: cannas@fisica.unipa.it

1. **INTRODUCTION**

The influence of ionizing radiation on optical properties of amorphous-$SiO_2$ (silica) is a timely research field strongly motivated by the large use of silica materials in many technologies requiring a good maintenance of transparency (e.g. fibers, laser optics and radioactive environments) [1, 2]. A huge number of works over the last decades have evidenced that irradiation induces intrinsic defects (dangling bonds, oxygen-deficiency or oxygen-excess) which cause optical transitions of absorption (OA) and photoluminescence (PL) (see review papers by Griscom [3] and Skuja et al. [4], and references therein). Owing to the complexity of OA and PL spectra over a wide range extending from visible to vacuum-ultraviolet (UV), the identification of defects and the assignment of the optical bands to their electronic structure remains a still incompletely solved problems.

One of the most controversial aspects regards the optical transitions associated with the paramagnetic non-bridging oxygen hole center (NBOHC) whose structure is identified by its paramagnetic properties and is denoted by $\equiv$Si-O• [5]; ($\equiv$) stands for bonds with three oxygen and (•) indicates un unpaired electron. It is well accepted that this defect gives rise to a weak OA band at 2.0 eV (oscillator strength $f_{2.0eV} \sim 1.5 \times 10^{-4}$) whose inverse transition is a PL at 1.9 eV with lifetime of $\sim$14 μs [6-8]. The main experimental proof that supports the assignment of these visible transitions to the NBOHC is the observation of a phonon side band of the zero-phonon-line of the 1.9 eV emission [6, 9], whose frequency (890 $cm^{-1}$) is very close to that of the Si-O stretching. The PL at 1.9 eV emission can be also excited in the UV range, its excitation (PLE) profile consisting of two large bands: one centered at 4.80 eV with full width at half maximum (FWHM) of 1.05 eV and the other centered at 6.4 eV with FWHM of 1.7 eV [10]. The identification of the corresponding UV absorption bands that correlate to the PL at 1.9 eV is limited to a few experiments [11, 12], also because of the overlap of bands associated with other defects which makes difficult the analysis of spectral shapes [3, 4, 13]. Recently, the optical properties of NBOHC have been also studied by computational works [14-16]; however, although the calculated optical transitions agree with the measured peak positions, the assignment of the energetic levels to molecular orbitals is controversial

In this paper we present experimental results concerning the two OA bands at 2.0 and 4.8 eV and the PL at 1.9 eV observed in a large number of wet synthetic silica samples with a different concentration of NBOHC's induced by γ- and β-ray irradiation over a wide dose range, five decades. Our purpose is to quantitatively examine the relation among the optical bands and extract spectroscopic parameters (oscillator strength and luminescence quantum yield) which supply useful information about the excitation/luminescence pathways occurring inside these defects.

## 2. EXPERIMENTAL METHODS

We investigated γ- and β-irradiated samples chosen among three representative synthetic wet silica specimens having a different OH content:

1. Suprasil 1 (S1), [OH]≈1000 ppm
2. Suprasil 311 (S311), [OH]≈200 ppm
3. Corning 7940 (CNG), [OH]≈800-1000

S1 and S311 materials were supplied by Heraeus [17], CNG was supplied by Corning [18].

The samples utilized in our experiments are slab shaped with square surface 5×5 mm$^2$ optically polished, thickness being $d$=1 mm and $d$=0.5 mm for the γ- and β-ray irradiation respectively. The γ irradiation was performed using a $^{60}$Co source (1.2 MeV), doses ranging from 2×10$^5$ to 10$^7$ Gy; β irradiation was performed by using a Van de Graff electron accelerator (2.5 MeV), doses ranging from 1.2×10$^6$ to 5×10$^9$ Gy.

OA measurements in the range 1.5-6.0 eV were made at room temperature by a double beam spectrometer (JASCO V-560), the bandwidth was 2 nm.. This instruments is equipped by a double monochromator on the excitation side which reduces the stray light down to 0.0003%.

Steady-state PL and PLE spectra were measured by a spectrofluoremeter (JASCO PF-6500), mounting a Xenon lamp of 150 W. The samples were mounted in the holder in the so-called 45° back-scattering geometry and a low-pass band filter was placed in front of the emission detector to eliminate the second order diffraction. Both for PL and excitation PL measurements the bandwidth was 5 nm. The excitation profiles were corrected for the spectral efficiency of the exciting light by using a Rhodamine B sample in glycerol as a reference. As regards the PL spectra, as they are plotted in intensity per energy interval units, they were corrected for the monochromator dispersion. In order to quantitatively compare emission spectra measured in different samples under visible and UV excitation, the relative PL amplitudes were corrected for the excitation intensity and the sample thickness. Moreover, in the heavily irradiated samples having a low UV transmission leads, the PL amplitude is corrected by the factor $\alpha d/(1-\exp(-\alpha d)$, $\alpha$ being the absorption coefficient, to keep a linear dependence of the PL amplitude on the concentration of luminescent centers [19].

3.  **RESULTS**

The effect of irradiation on the absorption of S1 silica is shown in Fig. 1 where is plotted the difference between the spectra of samples irradiated with β- ($5\times10^9$ Gy) and γ-rays ($2\times10^6$ Gy) and not irradiated.. The β irradiated sample shows a weak OA band in the visible peaked at 2.01±0.03 eV with FWHM of 0.44±0.04 eV and amplitude of 0.15±0.01 cm$^{-1}$, and a composite UV absorption were two contributions centered around 4.8 eV and 5.8 eV are distinguished. In the γ irradiated sample, the OA spectrum is qualitatively similar to previous one apart from the lower intensity; it exhibits a band at ~2.0 eV with amplitude of 0.009±0.002 cm$^{-1}$, its observation being almost at limit of our experimental uncertainty, and a composite UV absorption consisting of two contributions around 4.8 eV and 5.8 eV.

The link between the OA bands in the visible and UV range is clarified by looking at the PL spectra. Fig. 2 shows the emission observed in the γ-irradiated S1 sample under excitation at 2.3 eV and 4.8 eV respectively. The PL profiles bring close similarities as regards the emission peak centered at 1.93±0.01 eV regardless the excitation energy; the FWHM can be measured only under UV excitation and results 0.115±0.05 eV, since the overlap between with the 2.3 eV excitation does not allow to observe the complete emission shape. The PLE profile in the range 4.0-6.0 eV is also reported in the same figure and consist of a band peaked at 4.80±0.02 eV whose width at half maximum, measured on the lower energy side, is 0.53±0.03 eV.

The spectral parameters obtained from PLE spectra help for best fitting the UV absorption with two gaussian curves, the first with fixed peak energy (4.80 eV) and FWHM (1.05 eV), while the second, without constraints, is centered at 5.83±0.01 eV with FWHM of 0.85±0.02 eV and is associated with the E' centers, ≡Si• [19, 20].

OA and PL spectra measured in the other samples listed in section II show almost identical spectral features (position and FWHM) with those of Figs. 1 and 2 as concerns the absorption band at 2.0 eV and the PL at 1.9 eV. Moreover, by using the above described fit procedure to analyze the UV absorption spectra we single out the contribution of the 4.8 eV band. Fig. 3 shows the intensity or area of the 4.8 eV OA band, $I^{abs}_{4.8eV}$, against that of the 2.0 eV OA band, $I^{abs}_{2.0eV}$. The two bands are linearly correlated as evidenced by the best fit function, $y = ax$, the slope $a$=197±7 being the ratio between their intensities.

Finally, in Fig. 4 we report the dependence between the amplitudes of the PL at 1.9 eV, $A^{PL}_{4.8eV}$ and $A^{PL}_{2.3eV}$ excited at 4.8 eV and 2.3 eV, respectively. Data can be fitted by a linear correlation, $y = bx$, in which the coefficient $b$ is 93±4.

4. **DISCUSSION**

The above reported results point out the direct relationships between the OA bands at 2.0 and 4.8 eV and the PL at 1.9 eV excited under both OA bands. These findings are obtained on a wide number of synthetic wet silica samples, chosen among different materials and irradiated with different γ and β doses. Then, they are a clear experimental evidence that the overall optical activity arises from a single defect, the most accredited model being the NBOHC [4, 9].

Basing on the single-defect model, the correlation curves evidenced in Figs. 3 and 4 allow to extract quantitative parameters characterizing the optical transitions.

1) As the intensities of both OA bands, 2.0 eV and 4.8 eV, are proportional to the same defect concentration $N$ trough the expressions [21]:

$$Nf_{2.0eV} = K \cdot I^{abs}_{2.0eV} ; \tag{1.a}$$

$$Nf_{4.8eV} = K \cdot I^{abs}_{4.8eV} \tag{1.b}$$

with $K = \dfrac{n}{(n^2+2)^2} \times 9.111 \times 10^{15} \left[eV^{-1}cm^{-2}\right]$, $n$ being the refraction index, $I^{abs}_{4.8eV}/I^{abs}_{2.0eV}$ measures the ratio between the corresponding oscillator strengths, $f_{4.8eV}/f_{2.0eV}$. So, from the value $f_{2.0eV} \approx 1.5 \times 10^{-4}$, we get $f_{4.8eV} \approx 0.03$. We acknowledge that in a previous work $f_{4.8eV}$ was measured to be $\approx 0.05$, by the comparison between the intensities of the 2.0 eV and 4.8 eV band induced in a sample irradiated with $F_2$ (7.9 eV) laser photons at T=77 K [8].

2) The link between the PL at 1.9 eV and the two OA bands is accounted for the expressions [22]:

$$A^{PL}_{2..3eV} \propto I^{exc}(2.3eV) \eta_{2.0eV} \cdot \alpha(2.3eV)d ; \tag{2.a}$$

$$A^{PL}_{4.8eV} \propto I^{exc}(4.8eV) \eta_{4.8eV} \cdot \alpha(4.8eV)d \tag{2.b}$$

where $\eta_{2.0eV}$ and $\eta_{4.8eV}$ are the 1.9 eV luminescence quantum yields under visible and UV excitation, respectively. Since the PL amplitudes as measured in the present work are corrected for both $I^{exc}$ and $d$, the ratio between (2.a) and (2.b) gives:

$$\dfrac{A^{PL}_{2.3eV}}{A^{PL}_{4.8eV}} = \dfrac{\eta_{2.0eV}}{\eta_{4.8eV}} \cdot \dfrac{\alpha(2.3eV)}{\alpha(4.8eV)} \tag{3}$$

Hence, substituting the ratio between the PL under excitation at 4.8 and 2.3 eV, derived from Fig. 4, and that between the absorption coefficient at the same energies, we get:

$$\dfrac{\eta_{2.0eV}}{\eta_{4.8eV}} = 2.9 \pm 0.6$$

To account for the spectroscopic parameters related to the OA (2.0eV and 4.8eV) and PL (1.9eV) transitions, we sketch in Fig. 5 a three levels energetic diagram that was put forward to explain the optical properties of the NBOHC [6, 14-16]. In the ground state, the levels (*a*) and (*b*) are filled whereas the higher level (*c*) is partially occupied. In agreement with this scheme, we can distinguish between two different excitation/luminescence pathways: i) visible excitation (two-step process) in which the 2.0eV-OA ($f_{2.0eV}=1.5\times10^{-4}$) and the 1.9eV-PL are associated with the transitions between the levels (*b*) and (*c*); ii) UV excitation (three-step process) in which the 4.8eV-OA ($f_{4.8eV}=0.03$), occurring between the levels (*a*) and (*c*), is followed by a non radiative electronic relaxation from (*b*) to (*a*) that leaves a hole in (*b*) and finally, by the 1.9 eV radiative emission from (*c*) to (*b*).

As evidenced by our result, the three-step process (4.8-eVOA/1.9eV-PL) has a lower efficiency in comparison with the two-step process (2.0-eVOA/1.9eV-PL). We can interpret this finding on assuming that, after the OA transition at 4.8eV, the occurrence of the PL at 1.9 eV is governed by the competition between the non radiative relaxations (*b*)→(*a*), at a rate $k_{nr}^{b,a}$, and (*c*)→(*a*), at a rate $k_{nr}^{c,a}$. Since after the relaxation (*b*)→(*a*) the system is equivalent to that after excitation at 2.0 eV having quantum yield $\eta_{2.0eV}$, $\eta_{4.8eV}$ can be expressed by:

$$\eta_{4.8eV} = \frac{k_{nr}^{b,a}}{k_{nr}^{b,a} + k_{nr}^{c,a}} \cdot \eta_{2.0eV}. \qquad (4)$$

Hence, from the comparison with our finding, $\eta_{2.0eV}/\eta_{4.8eV} \approx 3$, we get $k_{nr}^{c,a} \approx 2 \cdot k_{nr}^{b,a}$.

We point out that, the two non radiative decay rates $k_{nr}^{b,a}$ and $k_{nr}^{c,a}$ are consistent with the hypothesis that the large Stokes shift between the OA at 4.8 eV and the PL at 1.9 eV is a consequence of a slow electronic relaxations down to a level, (*a*) in the Fig. 5, located below the top of the valence band [15, 16]. Moreover, as reported in Ref. [10], the 1.9 eV PL intensity decreases with temperature whereas its lifetime has a poor dependence on temperature. This indicates that the non radiative processes under UV excitation do not occur simultaneously with the radiative emission in agreement with the above reported scheme.

## 5. CONCLUSIONS

In summary, we studied the optical transitions (absorption at 2.0 eV and 4.8 eV, luminescence at 1.9eV) related to NBOHC induced in synthetic wet silica samples by $\gamma$ and $\beta$ irradiation. These bands keep a constant ratio regardless the sample, from which we determine: i) the absorption oscillator strengths value ($f_{2.0eV} \approx 1.5 \times 10^{-4}$, $f_{4.8eV} \approx 0.03$); ii) the comparison between the luminescence quantum yield under visible and UV excitation ($\eta_{2.0eV}/\eta_{4.8eV} \approx 3$). These results are consistent with a three-levels energetic scheme previously proposed in literature [6, 14-16], and account for the lower efficiency of the UV-excitation/luminescence pathway in which additional non radiative electronic relaxations take place.

**ACKNOWLEDGEMENTS**

The authors acknowledge the support and the helpful discussions with Prof. R. Boscaino and the research group people at University of Palermo; E. Calderaro for taking care of the $\gamma$ irradiation; and G. Napoli for technical assistance. This work was partially supported by a project (PRIN2002) of Italian Ministry of University and Research

**REFERENCES**

[1] *Defects in SiO$_2$ and Related Dielectrics: Science and Technology*, edited by G. Pacchioni, L. Skuja, and D.L. Griscom (Kluwer Academic Publishers, Dordrecht, 2000)

[2] *Structure and Imperfections in Amorphous and Crystalline Silicon Dioxide* edited by R.A.B. Devine, J-P. Duraud and E. Doory´ee (Wiley, Chichester, 2000)

[3] D.L. Griscom, J. Ceram. Soc. Japan **99**, 923 (1991).

[4] L. Skuja, M. Hirano, H. Hosono, and K. Kajihara, Phys. Stat. Sol. (c) **2**, 15 (2005)

[5] M. Stapelbroek, D. L. Griscom, E. J. Friebele, and G. H. Sigel Jr., J. Non-Cryst. Solids **32**, 313 (1979).

[6] L. Skuja, J. Non-Cryst. Solids **179**, 51 (1994).

[7] Y. Sakurai, J. Appl. Phys. **87**, 755 (2000)

[8] H. Hosono, K. Kajihara, T. Suzuki, Y. Ikuta, L. Skuja, and M. Hirano, Solid State Commun. **122**, 117 (2002).

[9] L. Skuja, T. Suzuki, and K. Tanimura, Phys. Rev. B **52**, 15208 (1995).

[10] M. Cannas, F.M. Gelardi, Phys. Rev. B **69**, 153201(2004).

[11] L. Skuja, K. Tanimura, and N. Itoh, J. Appl. Phys. **80**, 3518 (1996).

[12] M. Cannas and M. Leone, J. Non-Cryst. Solids **280**, 183 (2001).

[13] L. Skuja, M. Hirano, and H. Hosono, Phys. Rev. Lett. **84**, 302 (2000).

[14] T. Suzuki, L. Skuja, K. Kajihara, M. Hirano, T. Kamiya, and H. Hosono, Phys. Rev. Lett. **90**, 186404 (2003).

[15] T. Bakos, S. N. Rashkeev, and S.T. Pantelides, Phys. Rev. Lett. **91**, 226402 (2003).

[16] T. Bakos, S. N. Rashkeev, and S.T. Pantelides, Phys. Rev. B, **70**, 075203 (2004).

[17] Heraeus Quartzglas, Hanau, Germany, Catalogne POL-O/102/E.

[18] Corning Spa ISDP Europe, Tecnottica Consonni S.N.C., Calco (Lc), Italy.

[19] J. –R. Lakowicz, *Principles of Fluorescence Spectroscopy*, (Plenum, New York, 1983).

[19] R.A. Weeks and E. Sonder, in *Paramagnetic Resonance*, edited by W. Low (Academic Press, New York, 1963), pp. 869–879.

[20] R. Boscaino, M. Cannas, F.M. Gelardi, and M. Leone, Nucl. Instrum. Methods Phys. Res. B **116**, 373 (1996).

[21] A.M. Stoneham, *Theory of Defects in Solids*, (Clarendon, Oxford, 1975).

[22] S. Shinoya, in *Luminescence of Solids*, edited by D.R. Vij (Plenum Press, New York, 1998), pp. 95–133.

**FIGURE CAPTIONS**

**FIG. 1:** Difference absorption spectra between β- ($5\times10^9$ Gy) (a) and γ- ($2\times10^6$ Gy) (b) irradiated and non irradiated S1 samples. Dashed lines represent the two gaussian components that best fit the UV absorption.

**FIG. 2:** Normalized photoluminescence spectra detected under excitation at 4.8 eV (solid line) and 2.3 eV (empty symbol) in the S1 sample exposed to a $2\times10^6$ Gy γ-ray dose (left side); UV excitation spectrum monitored at 1.93 eV (right side).

**FIG. 3:** Correlation between the intensities of the 4.8 eV and 2.0 eV absorption bands measured in the S1, S311 and CNG irradiated samples. Solid line represents the linear best fitting.

**FIG. 4:** Correlation between the amplitudes of the 1.9 eV emission detected in S1, S311 and CNG irradiated samples under 2.3 eV and 4.8 eV excitation. Solid line represents the linear best fitting.

**FIG. 5:** Diagram of the energy levels and transitions accounting for the 1.9 eV PL excited at 2.0 eV and 4.8 eV. Dashed arrows represents the non radiative relaxations.

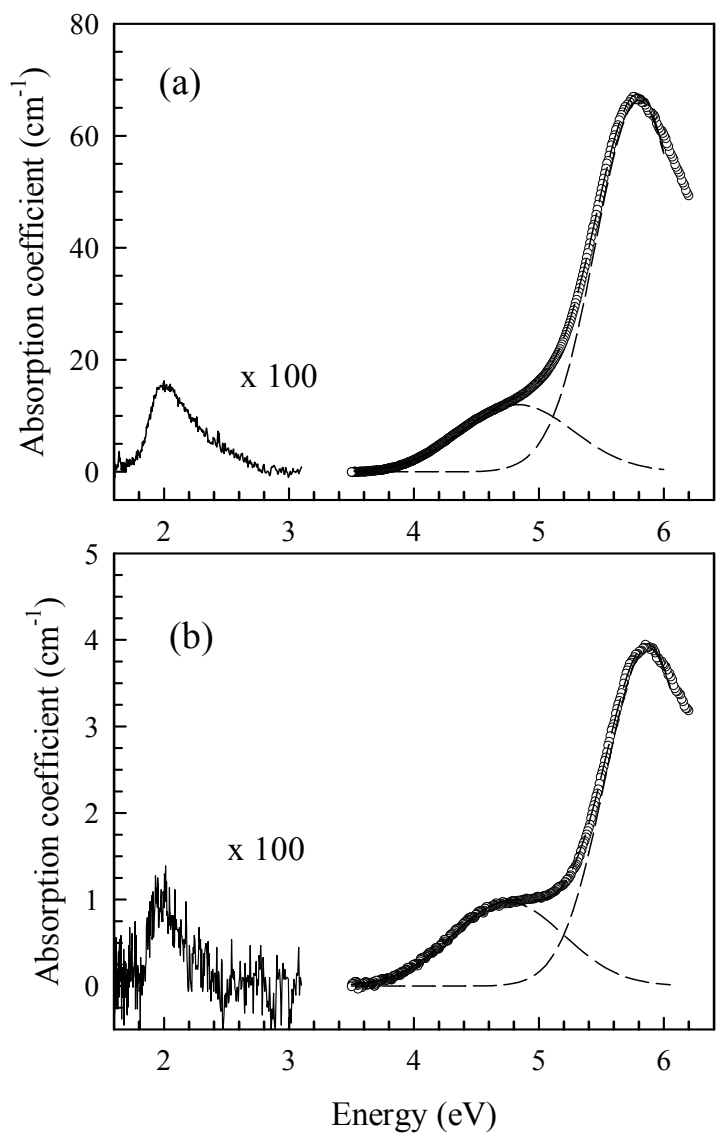

Figure 1

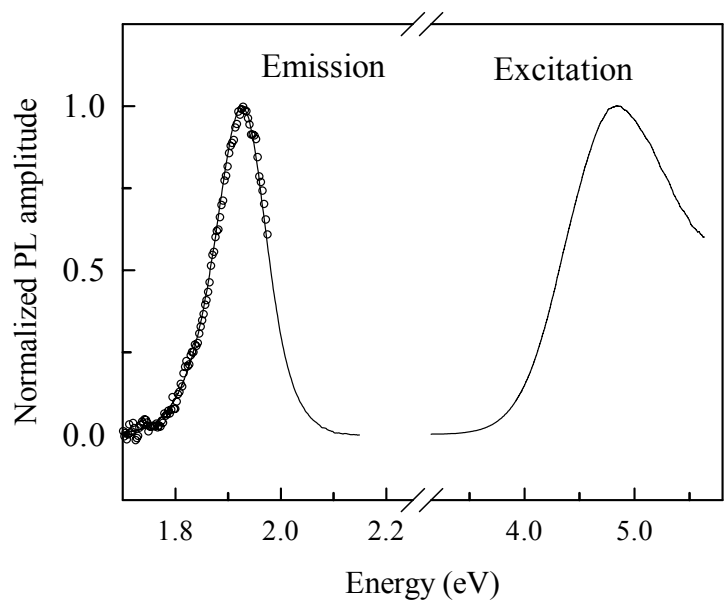

Figure 2

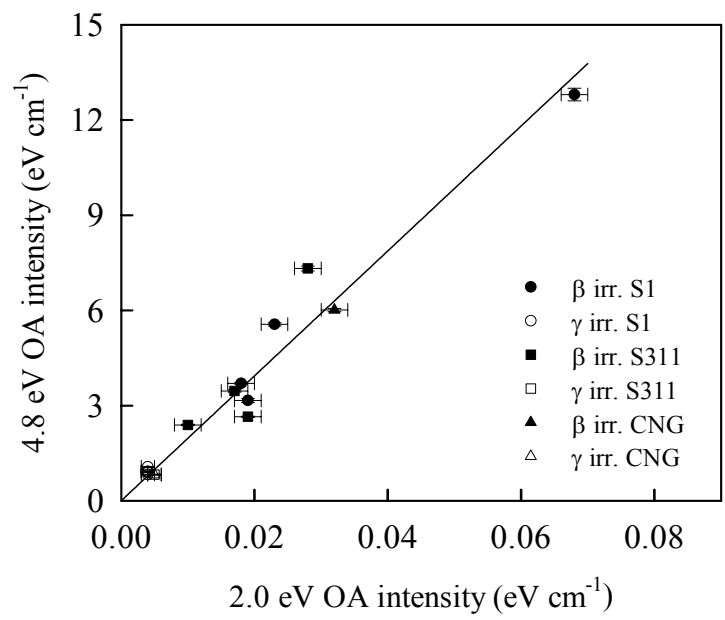

Figure 3

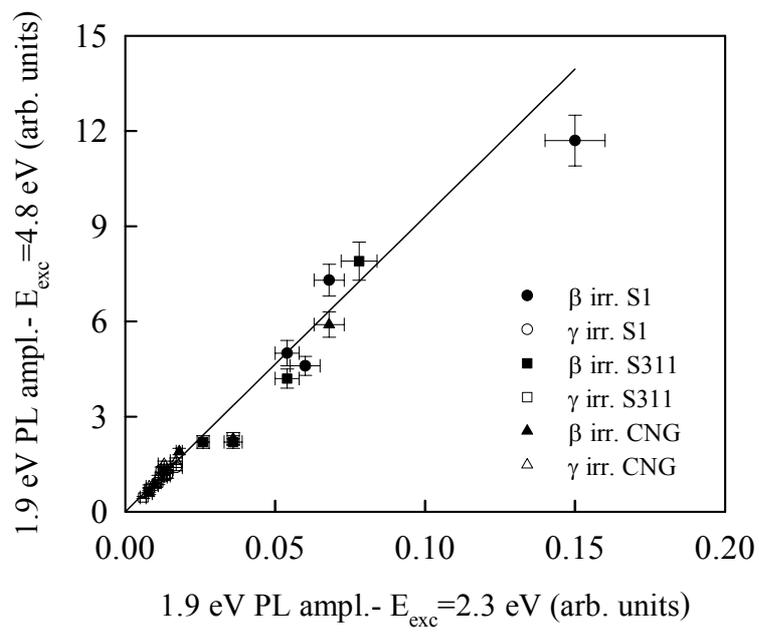

Figure 4

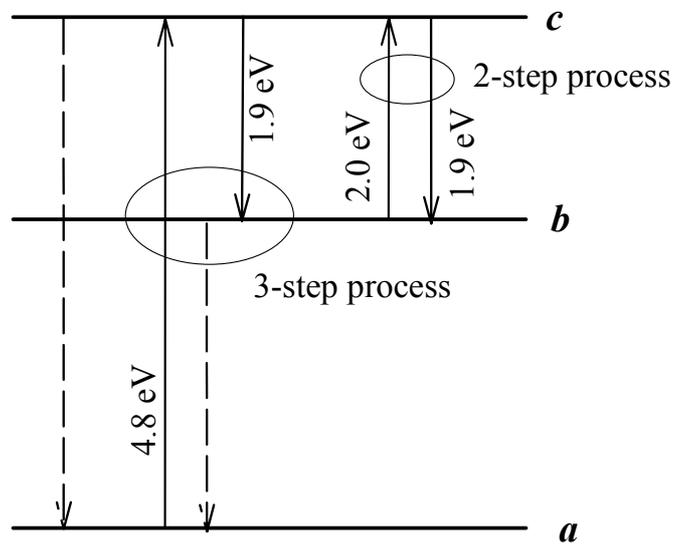

Figure 5